\input harvmac
\def \W {{\cal W}}

\def\vp{{\varpi}}
\def\a{\alpha}

\def \na {\nabla} 
\def \ep{\epsilon}
\def\D{\Delta}

\def \te {\textstyle}
\def \pa {\partial}

\def \const {{\rm const} }

\def \V {{\cal V}}
\def \F {{\cal F}}

\def \te {\textstyle }
\def \S {{\cal S}}

\def \V {{\cal V}}
 
\def \k {\kappa} 
\def \F {{\cal F}} \def \Te {{\rm T}}

\def \del {\partial}

\def \na {\nabla}
\def \const {{\rm const}}
\def \ha{{\textstyle{1\over 2}}}

\def \na {\nabla }
\def \D {\Delta}
\def \a {\alpha}
\def \b {\beta}
\def \ov {\over}
 
\def\r {\rho}
\def \p {\phi}
\def \m {\mu}
\def \n {\nu}
\def \vp {\varphi }
\def \t {\tau}
\def \td {\tilde }
\def \d {\delta}

\def \l {\lambda}

\def \P {\Phi}

\def \inv {^{-1}}
\def \ov {\over }
\def \four{{\textstyle{1\over 4}}}

\def \D {\Delta}
\def \d {\delta}

\def \lr { \lref}
\def \k {\kappa}

\def \four{{\textstyle{1\over 4}}}

\def \ep {\epsilon}
\def \p {\phi}

\def \del {\partial}
\def \vp {\varphi}
\def \hmn { h_{mn}}
\def \inv  {^{-1}}
\def \d {\delta} 
\def\a{\alpha} 
\def \b {\beta}
\def \G {\Gamma} 
\def \ov {\over} 
\def \eight { { \textstyle{ 1 \ov 8} } }
\def \sixt{ { \textstyle{ 1 \ov 16} } }
\def \C {\bar C} 
\def \CC {{\cal C} }
\def \trto { { \textstyle{ 1 \ov 32} } }

\def \T {{\cal T}^{(T)}}
\def \TT {{\cal T}^{(h)}}
\def \bz {\bar z}
\def \td {\tilde}

\def \half {{1 \ov 2}} 

\def\np {  {\it Nucl. Phys.} }
\def \pl { {\it  Phys. Lett.} }

\def \pr  { {\it Phys. Rev.} }

\lr \gibb{G.W. Gibbons and K. Maeda, 
``Black holes and membranes in higher-dimensional
theories with dilaton fields", 
{\it Nucl. Phys.} {\bf B298} (1988) 741;
C.G. Callan,Jr., S.S. Gubser,
I.R. Klebanov and A.A. Tseytlin, 
``Absorption of Fixed scalars and the D-brane Approach
to Black Holes", 
{\it Nucl. Phys.}  {\bf B489} (1997) 65, 
{\tt hep-th/9610172}. }
\lr\mbg{M. B. Green, ``Point-Like States for Type IIB Superstrings,''
\pl {\bf B329} (1994) 435, {\tt hep-th/9403040}. }
\lr \wilt{ D. Wiltshire, 
``Dilaton black holes with a cosmological
term", {\tt gr-qc/9502038}; 
S.J. Poletti, J. Twamley  and  D.L. Wiltshire,  
``Charged dilaton black holes with a cosmological
term", {\it  Phys. Rev.} {\bf D51} (1995) 5720, 
{\tt hep-th/9412076}; K.C.K. Chan, J.H. Horne
and R.B. Mann, ``Charged dilaton black holes with unusual asymptotics", 
{\it Nucl. Phys.} {\bf  B447} (1995) 441, {\tt  gr-qc/9502042.}}

\lr \DH { L. Dixon and J. Harvey, ``String theories in ten dimesnions without space-time supersymmetry",  
{\it Nucl. Phys.} {\bf B274} (1986) 93;
 N. Seiberg and E. Witten,
``Spin structures in string theory", 
{\it Nucl. Phys.} {\bf B276} (1986) 272;
C. Thorn, unpublished.}

\lr\Sasha{A.M. Polyakov, ``String theory and quark confinement,''
{\it Nucl. Phys. B (Proc. Suppl.)} {\bf 68} (1998) 1, {{\tt
hep-th/9711002}}. }

\lr\berg{O. Bergman and M. Gaberdiel, ``A Non-supersymmetric Open
String Theory and S-Duality,'' \np {\bf B499} (1997) 183,
{\tt hep-th/9701137}. }

\lr  \kleb{
I.R. Klebanov, ``World volume approach to absorption by nondilatonic branes,''
  {\it Nucl. Phys.} {\bf B496} (1997) 231,
  {{\tt hep-th/9702076}}; 
S.S. Gubser, I.R. Klebanov, and A.A. Tseytlin, ``String theory and classical
 absorption by three-branes,'' {\it Nucl. Phys.} {\bf B499} (1997) 217,
  {{\tt hep-th/9703040}}.}

\lr  \gkThree{
S.S. Gubser and I.R. Klebanov, ``Absorption by branes and Schwinger terms in
  the world volume theory,'' {\it Phys. Lett.} {\bf B413} (1997) 41,
  {{\tt hep-th/9708005}}.}

\lr  \jthroat{
J.~Maldacena, ``The Large N limit of superconformal field theories and
  supergravity,'' {\it Adv. Theor. Math. Phys.} {\bf 2} (1998) 231, 
{{\tt
  hep-th/9711200}}.}

\lr  \US{
S.S. Gubser, I.R. Klebanov, and A.M. Polyakov, ``Gauge theory correlators
  from noncritical string theory,'' {\it Phys. Lett.} {\bf B428} (1998)
105,
  {{\tt hep-th/9802109}}.}

\lr  \EW{
E.~Witten, ``Anti-de Sitter space and holography,''
 {\it Adv. Theor. Math. Phys.} {\bf 2} (1998) 253, 
 {{\tt hep-th/9802150}}.}

\lr  \Ed{
E.~Witten, ``Anti-de Sitter Space, Thermal
Phase Transition, and Confinement in Gauge Theories,''
{\it Adv. Theor. Math. Phys.} {\bf 2} (1998) 505, 
{\tt hep-th/9803131.}}

 \lr  \gkp{
S.S. Gubser, I.R. Klebanov, and A.W. Peet, ``Entropy and temperature of
  black 3-branes,'' {\it Phys. Rev.} {\bf D54} (1996) 3915,
  {{\tt hep-th/9602135}}.}

\lr  \KT{
I.R. Klebanov and A.A. Tseytlin, 
``Entropy of near-extremal black p-branes,"
\np {\bf B475} (1996) 164, 
{\tt hep-th/9604089.}} 

\lr\GO{D.J. Gross and H. Ooguri,
``Aspects of Large N Gauge Theory Dynamics as Seen by String Theory,''
\pr {\bf D58} (1998) 106002, {\tt hep-th/9805129}.}

\lr  \KS{
I.R. Klebanov and L. Susskind,
``Schwarzschild Black Holes in Various Dimensions from Matrix
Theory,'' {\it Phys. Lett.} {\bf B416} (1997) 62,
{\tt hep-th/9709108.}}

\lr  \Itz{
N. Itzhaki, J. Maldacena, C. Sonnenschein, S. Yankielowicz,
``Supergravity and the Large N Limit of Theories with 16 Supercharges,''
{\it Phys. Rev.} {\bf D58} (1998) 046004,
{\tt hep-th/9802042.}}

\lr  \HP{
S.W. Hawking and D. Page,
``Thermodynamics of black holes in anti de Sitter space,"  
{\it Commun. Math. Phys.} {\bf 87 } (1983) 577.}

\lr  \ooguri{
C. Csaki, H. Ooguri, Y. Oz and J. Terning,
``Glueball Mass Spectrum From Supergravity,''
{\tt hep-th/9806021.}}

\lr  \jev{
R. De Mello Koch, A. Jevicki, M. Mihailescu and  J. Nunes,
``Evaluation Of Glueball Masses From Supergravity,''
{\tt hep-th/9806125.}}

\lr  \aki{
A. Hashimoto   and Y. Oz, ``Aspects of QCD Dynamics from String Theory,''
{\tt hep-th/9809106.}}

\lr  \AP{
A.M. Polyakov, ``The Wall of the Cave,''
{\tt hep-th/9809057.}}

\lr  \brane{
J.~Polchinski, ``Dirichlet Branes and Ramond-Ramond charges,'' {\it Phys. Rev.
  Lett.} {\bf 75} (1995) 4724,
{{\tt hep-th/9510017}}. }

\lr  \Jbook{
J. Polchinski, ``String Theory,'' vol. 2, Cambridge University Press,
1998.}

\lr\rota{C. Csaki, Y. Oz, J.
 Russo  and J. 
      Terning, ``Large N QCD from Rotating Branes", 
{\tt hep-th/9810186}.}

      \lr\TASI{
J. Polchinski, ``TASI Lectures on D-Branes,''
{\tt hep-th/9611050.}  }

\lr  \Witten{
E.~Witten, ``Bound states of strings and p-branes,'' {\it Nucl. Phys.} {\bf
  B460} (1996) 335, {{\tt
  hep-th/9510135}}.  }

\lr  \hsdgt{
G.T. Horowitz and A.~Strominger, ``Black strings and p-branes,'' {\it Nucl.
  Phys.} {\bf B360} (1991) 197;
M.J. Duff  and J.X. Lu, 
``The selfdual  type IIB  superthreebrane,"
{\it Phys. Lett.}  {\bf B273} (1991)  409;
 G.W. Gibbons and  P.K. Townsend,
``Vacuum interpolation in supergravity via super p-branes",
 {\it Phys. Rev. Lett.} {\bf 71} (1993) 3754, 
{\tt hep-th/9307049}.}

\lref \tat {A.A. Tseytlin, 
``On the tachyonic terms in the string effective action", 
{\it Phys Lett.}  {\bf B264} (1991) 311.}
\lref \bank{T. Banks, ``The tachyon potential in string theory",
{\it Nucl. Phys.} {\bf B361} (1991) 166.}

\lref\banksus{T. Banks and L. Susskind, ``Brane--anti-brane   forces",
{\tt hep-th/9511194}.}

\lr\tva{
A.A.  Tseytlin,
``Cosmological solutions with dilaton and maximally
symmetric space in string theory", 
{\it Int. J. Mod. Phys.} {\bf D1} (1992) 223, 
 {\tt hep-th/9203033.}  } 
  \lr\bisa{
M. Bianchi and  A. Sagnotti,
``On the Systematics of Open String Theories",
{\it Phys. Lett.} {\bf B247}  (1990) 517.}
\lr\sagn{A. Sagnotti, ``Some Properties of Open - String Theories", 
{\tt hep-th/9509080}; ``Surprises in Open-String Perturbation Theory", 
 {\it Nucl.Phys.Proc.Suppl.} {\bf  B56}  (1997) 332, 
{\tt hep-th/9702093}; 
C. Angelantonj, ``Nontachyonic Open Descendants of the 0B String Theory",
{\tt  hep-th/9810214}. 
}
\lr\mig{A.A. Migdal, ``Hidden Symmetries of Large N QCD,"
{\it Prog. Theor. Phys. Suppl.} {\bf 131} (1998) 269, {\tt
hep-th/9610126}.}
\lr\alva{E. Alvarez, C. Gomez and T. Ortin,
{\tt hep-th/9806075}. }

\baselineskip8pt
\Title{\vbox
{\baselineskip 6pt
{\hbox {PUPT-1819}}{\hbox{Imperial/TP/98-99/07  }}
{\hbox{hep-th/9811035}} 
{\hbox{   }}
}}
{\vbox{\vskip -30 true pt
\centerline {D-Branes and Dual Gauge Theories}
\medskip
\centerline {in Type 0 Strings }
\medskip
\vskip4pt }}
\vskip -20 true pt 
\centerline{ Igor R. Klebanov}
\smallskip\smallskip
\centerline{Joseph Henry Laboratories, Princeton University, 
Princeton, New Jersey 08544, USA}
\bigskip
\centerline  {Arkady A. Tseytlin\footnote{$^{\dagger}$}{\baselineskip8pt
Also at  Lebedev  Physics
Institute, Moscow.} }
\smallskip\smallskip
\centerline {  Blackett Laboratory, Imperial College,  London SW7 2BZ, U.K.} 

\bigskip\bigskip
\centerline {\bf Abstract}
\baselineskip10pt
\noindent
\medskip
We consider the type $0$ theories, obtained from 
the closed NSR string  by a diagonal GSO
projection which excludes space-time fermions,
and study the D-branes in these theories. The
low-energy dynamics of $N$ coincident D-branes is governed by a 
$U(N)$ gauge theory coupled to adjoint scalar fields.
It is tempting to look for the type $0$ string
duals of such bosonic gauge theories
in the background of the R-R charged
$p$-brane classical solutions. This results in a picture analogous to
the one recently proposed by Polyakov (hep-th/9809057). 
One of the serious problems that
needs to be resolved is the closed string tachyon mode which couples
to the D-branes and appears to cause an instability. We study the tachyon
terms in the type $0$ effective action and argue that the background
R-R flux provides a positive shift 
of the (mass)$^2$ of the tachyon. Thus, for sufficiently large
flux, the tachyonic instability may be cured, removing
the most basic obstacle to constructing the type $0$ duals of
non-supersymmetric gauge theories. We further find that the tachyon
acquires an expectation value in presence of the R-R flux.
This effect is crucial for breaking the conformal invariance in the
dual description of the $3+1$ dimensional non-supersymmetric gauge
theory.

\bigskip
 
\Date {November 1998}


\noblackbox \baselineskip 15pt plus 2pt minus 2pt 

\centerline{\ \ \ \ \ \ \ \ \ \ \ \ \ \ \  \ \ \ \it ``Presumably this tachyon should ... peacefully condense in the bulk" \AP}

\newsec{Introduction}

The idea that supergravity may be used to obtain exact results
in strongly coupled ${\cal N}=4$ SYM theory with a large number of
colors \refs{\kleb,\gkThree,\jthroat,\US,\EW} has been explored 
extensively in recent literature. Of course, a more ambitious goal
is to extend this success to a stringy description of 
non-supersymmetric Yang-Mills theory in $d$ dimensions.
One fruitful approach to this problem, proposed in \refs{\Ed},
is to start with a maximally supersymmetric gauge theory in $d+1$
dimensions and heat it up, so that the temperature breaks all
supersymmetry. A supergravity description of such thermal
strongly coupled gauge theories is indeed possible in
terms of near-extremal brane solutions \refs{\gkp,\KT,\KS,\Itz},
whose near-horizon geometry corresponds to a black hole in AdS
space \refs{\Ed,\HP}.
The new insights into non-supersymmetric gauge theories obtained
in \refs{\Ed}, and further developed in \GO, 
include a new conceptual approach to the quark confinement
and to the discreteness of the glueball spectrum.
Subsequent work has revealed an interesting structure of the
glueball spectra calculated in a limit akin to the lattice
strong coupling limit \refs{\ooguri,\jev,\aki,\rota}. However,
extrapolation of these calculations to the continuum (weak coupling)
limit poses a very difficult problem because it corresponds to the
large $\alpha'$ limit of string theory.

A rather different approach to non-supersymmetric gauge theories was
suggested in \refs{\AP}, building on earlier results
in \Sasha. It is based on a non-supersymmetric 
string theory  obtained via a diagonal (non-chiral) GSO projection
\DH\ which is usually referred to as
type 0 (A or B) theory  \refs{\Jbook} (the type is taken to
count the number of supersymmetries).\foot{In \refs{\mig,\alva} it was
also suggested that strings with world sheet supersymmetry may be
related to non-supersymmetric gauge theories.}
In  $D=10$ the spectra of these theories are:\foot{
We adopt the notation of \refs{\Jbook}: for example,
$NS-$ and $NS+$ refer to odd
or even chiral world sheet fermion number
in the NS sector.  }

\noindent
$
\ \ \ \ \ \ \ \ {\rm type} \ 0A : \ \ \ 
(NS-,NS-)\oplus (NS+,NS+)   \oplus(R+,R-) \oplus(R-,R+)\ ,$

\noindent  $
\ \ \ \ \ \  \ \ {\rm type} \ 0B: \  \ \ (NS-,NS-)\oplus(NS+,NS+) 
\oplus(R+,R+) \oplus(R-,R-)\ .$

\noindent
Both of these theories have no fermions in their spectra
but produce modular invariant
partition functions \refs{\DH,\Jbook}.\foot{
One may ask, for example, why we cannot have
a consistent theory where only the  
$(NS+,NS+) \oplus(R+,R+)$ part of the full type IIB spectrum (which is
$(NS+,NS+) \oplus(R+,R+) \oplus(NS+,R+)\oplus(R+,NS+)$) is kept.
The answer is that such a theory would not be modular invariant.
 }
The massless bosonic fields are  as in the corresponding 
type II theory (A or B), 
but with the  doubled set of the Ramond-Ramond (R-R)  fields. 
One of the main claims of
\refs{\AP} is that, although the spectrum of 
type 0 theory contains a tachyon from the $(NS-,NS-)$ sector,
this theory can be made to describe a tachyon-free gauge
 theory.\foot{Let us note that type 0B theory  has open  string
 descendants
which were originally constructed  by orientifold projection 
in \bisa. They, in general, have open-string tachyons in their spectra. 
A non-tachyonic   model  
(which is anomaly-free and contains chiral fermions in its 
open-string  spectrum)
was  found 
by a special Klein-bottle projection of
the 0B theory in \sagn.}

In this paper we 
further study possible relations between the
type $0$ strings and non-supersymmetric gauge theories (for now
we restrict ourselves to the critical dimension $D=10$, postponing
the non-critical case discussed in \refs{\AP} for the future).
We follow a route analogous to the one that has proven successful
in the context of type II theory and construct gauge theories by
stacking large numbers of coincident D-branes \refs{\brane,\Witten}. 
D-branes certainly can be introduced into the type $0$ string theory,
and the only question that needs to be addressed is: what is the
appropriate GSO projection for {\it open} strings attached to the D-branes?
\foot{Our discussion of the D-branes has significant
overlap with section 3.1 and Appendix C of an earlier paper \berg, 
which we were unfortunately unaware of at the time of writing, 
where the D-branes of type 0B theory and their
boundary states were discussed. 
We thank O. Bergman for pointing out this omission from the original
version of this article.}

Since the type $0$ theory has no space-time fermions in the bulk, it is
fairly obvious that there are none localized on a single D-brane
either, i.e. the Ramond sector of the open strings has to be removed.
In the Neveu-Schwarz open string sector we adopt the same projection
as for the bulk NS-NS states and keep only the vertex operators of
even fermion number in the $0$-picture.
This way the tachyon of the NS sector is projected out, and
the low-energy degrees of freedom on a D$p$-brane consist of
the massless gauge fields and scalars only.
We will see that this conclusion continues to hold for parallel
like-charged D-branes where the gauge theory naturally generalizes
to $U(N)$.
Thus, there is {\it  no tachyon}
on parallel D-branes  in type 0 theory.
This is in accord with the
conclusion reached in  \AP\ that the dual gauge theory
has no tachyonic states. 
The resulting D-brane action in the quadratic
approximation may be obtained by dimensionally reducing 
$9+1$ dimensional pure glue theory to $p+1$ dimensions.
This purely bosonic $U(N)$ gauge theory may therefore be built
by stacking $N$ coincident D$p$-branes in the type 0 
(0A for $p$ even, 0B for $p$ odd) theory, analogously
to the construction of maximally supersymmetric gauge theories by stacking
the D-branes of the type II theory. 
By analogy, we anticipate that the dual (closed string) description
of the gauge theory on a large number $N$ of
D$p$-branes involves the
classical R-R-charged $p$-brane background. Here, however, we
encounter a puzzle because the theory in the bulk appears to
suffer from a tachyonic instability. Since no such instability
is seen in the gauge theory on the D-branes, the conjectured
duality to a closed string background suggests that there 
the tachyonic instability disappears due to some non-perturbative
mechanism. 

With this idea in mind, we have
examined the structure of the low-energy effective
action of type 0 theory, with the emphasis on terms
involving the tachyon field $T$. We find that their structure in
the type $0$ theory is 
far more constrained than in the conventional bosonic string.
For example, due to the effects of the world sheet fermions in the
vertex operators, it is immediately clear that the
graviton-dilaton-tachyon part of the effective action has no {\it odd}
 powers
of the tachyon field. This implies that the tachyon effective potential
has no $T^3$ term, so that a positive $T^4$ term may cure the
instability. While this raises the possibility of the 
tachyon condensation in the absence of the R-R backgrounds, 
we find even more intriguing results 
for the terms that couple   the R-R $n$-form gauge field
strength, $F_{\mu_1 \ldots
\mu_n}$, to the tachyon: we demonstrate 
the presence of $F^2 T^2$ terms which shift the effective (mass)$^2$
of the tachyon field. Moreover, the sign of such terms
is such that the effective (mass)$^2$ can become positive
in the presence of R-R flux. 
Since the conjectured type $0$ duals of the gauge theories
necessarily include the R-R flux, this gives an appealing mechanism for
stabilizing such backgrounds.

While our results amount to a scenario for generalizing
the AdS/CFT duality to type $0$ string backgrounds, we should
caution the reader that this scenario is hard to establish
unambiguously based on
on-shell amplitude calculations, which is the only tool
we have used so far. Thus, for instance, the $F^2 T^2$ term
in the effective action 
is difficult to distinguish from a derivative coupling
$F^2 T \nabla^2 T$. Nevertheless, our analysis of the effective
action does make a rather convincing case for the stabilization
of the tachyon  in the presence of R-R charges.  
We emphasize that the mechanism for
stabilization is inherent to the type $0$ string, which is based on
supersymmetric world sheet theory, and would not be available in
the conventional bosonic string. 
This is related to the fact that in the type 0 string,
  but not in the conventional bosonic string,
the D-branes carry R-R charges. In this respect the
type $0$ theory is much more similar to the type II theory 
than to the bosonic string.

The structure of the paper is as follows. In section 2 we introduce the
D-branes of type $0$ theory and study their interactions and world volume
effective actions. In section 3 we calculate various three- and
four-point functions, and deduce the corresponding terms in the bulk
effective action. In section 4 we show that the near-horizon region
of the type $0$ threebrane solution is {\it not} of the form
$AdS_5\times S^5$, in agreement with the fact that the dual 
non-supersymmetric gauge theory is not conformal.
In Appendix A we further discuss the bulk effective action by calculating
the two graviton -- two tachyon  amplitude.
In Appendix B we  present the  general form of the 
classical  equations for the 
electric threebrane background of type 0 theory.

\newsec{Forces between parallel D-branes}
In this section we calculate the interaction energy of two parallel
D-branes in the type $0$ theory. This will provide a useful check of
our prescription for the GSO projection in the open string sector.
The calculation of the cylinder amplitude is analogous to the
corresponding calculation in type II theory \refs{\brane,\mbg}. 
In fact, the only
difference is that now we omit the contribution of the open string
$R$ sector. Thus, we find
\eqn\amp{
A= V_{p+1}\int_0^\infty {dt\over 2t}
(8\pi^2\alpha' t)^{-(p+1)/2} e^{-{t Y^2\over 2\pi\alpha'}}
\left ( \left [{f_3(q)\ov f_1(q)}\right ]^8 - 
\left[{f_4(q)\ov f_1(q)}\right ]^8
\right )\ ,
}       
where $q= e^{-\pi t}$ and we adopt the notation of \refs{\TASI}.
The first term in parenthesis comes from the NS spin structure, while
the second from the NS$(-1)^F$ spin structure. Just as in the type II
case, the negative power of $q$ corresponding to the open string
tachyon cancels between these two terms.

In order to factorize the cylinder in the closed string channel,
it is necessary to expand the integrand for small $t$.
For the NS-NS closed string exchange, which corresponds to the 
NS open string spin structure, we have
\eqn\yy{
 \left [{f_3(q)\ov f_1(q)}\right ]^8= t^4 ( e^{\pi/t} + 8 + O(e^{-\pi/t}) )
\ . } 
The first term corresponds to the NS-NS tachyon with $m^2 = -2/\a'$, 
while the second to the
attractive massless exchange. This indicates that a D-brane serves
as a source for the tachyon, as well as for the graviton and dilaton.

For the R-R closed string exchange, 
which corresponds to the 
NS$(-1)^F$ open string spin structure, we have
\eqn\ee{
- \left [{f_4(q)\ov f_1(q)}\right ]^8  = t^4 ( -16 + O(e^{-2\pi/t}) )
\ . }
Here, as expected, we do not find the tachyon.

Note that, while in the type II case the R-R repulsion exactly
balanced the NS-NS attraction, here  
the R-R massless repulsion is {\it twice}  as strong as the NS-NS massless
attraction. We believe that 
this is due to the doubling of the number of R-R
fields in the type 0 theory. For instance, while in the
type IIB case we have only the $(R+,R+)$ states coming from the
positive chirality spinors, in the type 0B theory we also have
the $(R-, R-)$ states coming from the negative chirality spinors.
The spectrum of the 0B theory thus contains two R-R scalars, 
two R-R 2-forms, and also a 4-form without the constraint that 
its 5-form field strength is selfdual.

A question that comes to mind immediately is whether the two parallel
D-branes will actually fly apart when the tachyon is removed
from the spectrum (this is not the conclusion that we want to reach,
since we want to build a gauge theory by stacking the D-branes on top
of each other). A possible way to evade this conclusion is to note
that the tachyon will work to attract the
D-branes to a sufficiently small distance. 
Only when a large number of D-branes have coalesced to
form a bound state with a macroscopic R-R field around it,  will the
tachyon (mass)$^2$ become shifted to remove the instability.
Unfortunately,  we cannot make this argument quantitative at present.


On the basis of the cylinder calculation and the low-lying spectra of
the open and closed strings, it is not hard to write down the 
low-energy action for a D-brane. It has the
Born-Infeld  form  similar to the
analogous action in type II theory except that now the R-R spectrum is
doubled, and there is also a tachyon in the NS-NS sector.
Thus, we find
$$S_p = - \Te_p \int d^{p+1} \sigma\  e^{-\Phi (X)} \sqrt {- \hat G}
\bigg[ 1+ \ \four
q \bar q\ T(X) \  + \four  \F_{\alpha\beta}^2 \ + \ha 
\sum_{i=1}^{9-p} (\partial_\alpha X^i)^2 + \ldots \bigg]
$$
\eqn\dact{ +\  \mu_p \int d^{p+1} 
\sigma \ \bigg( q C_{p+1} + \bar q \bar C_{p+1} \bigg)
\ ,
}
where $T$ and $\Phi$ are the background (bulk)  
tachyon and dilaton fields, 
$\hat G_{\alpha\beta}$ is the metric induced on the 
D-brane, and $C_{p+1}$ and $\bar C_{p+1}$ are the projections 
of the R-R fields.\foot{Here we use
the normalization of $C_{p+1}$ (and $\bar C_{p+1}$) as in  \TASI\ 
(i.e. $L= -{ 1 \ov 2 \k^2} R  + \ha |F_{p+2}|^2$, 
$|F_{p+2}|^2 = { 1 \ov (p+2)!} F_{p+2} F_{p+2}$). The  tachyon 
is normalized so that 
$L(T) ={ 1 \ov 8 \k^2} (\del^n T \del_n T  + m^2 T^2), 
\ \ m^2 = - {2 \ov \a'}$. 
   We  used the notation $\F_{\a\b} \equiv  2\pi \a' F_{\a\b} + B_{\a\b} $
where $F_{\a\b}$ is 
the world-volume vector  field strength 
 and 
absorbed $2\pi$ into the collective coordinates $X^i$.
We also  omitted other possible 
$\F\wedge ...\wedge  \F\wedge (C_k + \bar C_k)$ terms.} 

The discrete charges $q=\pm 1$ and $\bar q=\pm 1$ distinguish the
branes with respect to $C_{p+1}$ and $\bar C_{p+1}$
from the anti-branes.\foot{The presence of the factor
$q \bar q$ in the coupling of the D-brane to the tachyon
is related to the effective action R-R couplings $F\bar F T$
whose existence we demonstrate in section 3.1.}
The cylinder amplitude in \amp\ gives the potential between the
like-charged branes: $q_1 = q_2$, $\bar q_1 = \bar q_2$.
In order to study the case $q_1 =- q_2$, $\bar q_1 =- \bar q_2$
we need to change the sign of the NS$(-1)^F$ term which corresponds to
the R-R interaction. Then, just as in the type IIB case 
\refs{\mbg,\banksus}, 
the open string tachyon is no longer projected out.

Finally, we should be able to construct the cylinder amplitude for
$q_1 =- q_2$, $\bar q_1 = \bar q_2$ or
$q_1 = q_2$, $\bar q_1 =- \bar q_2$. Then the R-R contribution 
to the interaction potential  cancels
between the $C_{p+1}$ and $\bar C_{p+1}$ exchanges. 
The NS-NS couplings of the two D-branes are also different
now due to the factor $q\bar q$ in the source term for the
tachyon. The correct projection in the open string channel 
is to retain the R sector only \berg, so that the
the cylinder amplitude becomes
\eqn\newamp{
A= -V_{p+1}\int_0^\infty {dt\over 2t}
(8\pi^2\alpha' t)^{-(p+1)/2} e^{-{t Y^2\over 2\pi\alpha'}}  
\left [{f_2(q)\ov f_1(q)}\right ]^8
\ .
}       
Expanding for small $t$, we have
\eqn\yy{
- \left [{f_2(q)\ov f_1(q)}\right ]^8= t^4 ( - e^{\pi/t} + 8 + O(e^{-\pi/t}) )
\ , } 
so that the sign of the closed-string tachyon contribution is
indeed reversed compared to \amp.
The open string tachyon does not appear in this loop diagram,
but, somewhat unexpectedly, the fermions do.
Thus, the worldvolume theory of {\it two}  D-branes with
$q_1 =- q_2$, $\bar q_1 = \bar q_2$ or
$q_1 = q_2$, $\bar q_1 =- \bar q_2$ contains fermions in addition
to bosons. 
Since the total charge $q$ or $\bar q$ vanishes for such a composite
D-brane, it decouples from the bulk tachyon.

Altogether we find 4 different types of `elementary'
D-branes depending on the signs of $q$ and $\bar q$ (i.e.
there are 2 kinds of D-branes, and each one has a corresponding
antibrane). An alternative
way of labeling the D-branes and the R-R fields is to form
\eqn\comb{ q_{\pm} = \ha ({q \pm \bar q}) \ ,
\qquad  (C_{p+1})_{\pm} = {\textstyle { 1 \ov \sqrt 2}}
(C_{p+1} \pm \bar C_{p+1}) 
\ .
}
The D-branes with $q_-=0$ are sources of $C_+$ only, while
the D-branes with $q_+=0$ are sources of $C_-$ only. 
Note that for $p=3$ the cases $q_+=1, q_-=0$  ($q_+=0, q_-=1$)
correspond to electrically  (magnetically) charged
D3-branes. The presence of the two types of 3-branes is related
to the fact that in type 0B theory the 5-form field
strength is unconstrained (in type IIB theory it satisfies the
selfduality constraint which forces the 3-branes to be
dyonic). For example, for an electric 3-brane the
last term of \dact\ is replaced by
$$ \mu_5 \int d^4 \sigma\  C_4\ ,
$$
where $F_5 = dC_4$ is the unconstrained 5-form field strength.
To construct the self-dual 3-branes of type 0 theory we
need to bring together equal numbers of 
the elementary electric and magnetic 3-branes. 
In this paper we will mainly 
restrict ourselves to discussing purely electric or
purely magnetic configurations.

In the type II case the D-brane tensions and 
charge densities were normalized by Polchinski
on the basis of the cylinder calculation, and we will adopt the
same method. Since in \amp\ we have discarded the contribution of the
open string R-sector, the graviton-dilaton attraction comes from the
NS sector alone and turns out to be $1/2$ of what it was in the type
II case. This implies that the tension is reduced by a
factor of $\sqrt 2$, i.e. $ \Te_p = {\Te_p^{\rm II}/\sqrt 2}$.
However, the relation 
between  tension $\Te_p$  and  charge density $\m_p$  remains as
 it was in type II theory \TASI\ 
\eqn\cha{ \Te_p = {\mu_p\over \sqrt 2 \kappa }\ ,
}
where $\kappa$ is the gravitational constant.
Thus 
\eqn\mmm{
\mu_p = {\mu_p^{\rm II}\ov \sqrt 2} = \sqrt \pi 
\left (2\pi \sqrt {\alpha'} \right )^{3-p}
\ .
}
This assignment ensures that the massless
R-R exchange term is also normalized correctly: both the $C_{p+1}$
and the $\bar C_{p+1}$ contributions work out to be $1/2$ of the
type II result, so that the sum is equal to the type II result.
Indeed, the R-R exchange term, which comes from the NS$(-1)^F$
sector, is identical in the type $0$ and type II theories.

A similar argument shows that the Dirac-Nepomechie-Teitelboim charge
quantization condition is satisfied by the type $0$
D-branes. The net phase $\theta$
accumulated in circling a $p$-brane around the
$(6-p)$-brane magnetically dual to it, 
is doubled due to the presence of both the $C_{p+1}$ and
$\bar C_{p+1}$ fields:
$$ \theta = 2 \mu_p \mu_{6-p} = 2\pi
\ .$$
This shows that, just as in the type II theory, D-branes carry minimal
charges consistent with the quantization condition.


\newsec{ Tree-level amplitudes and the effective action}

In this section we perform a number of on-shell amplitude
calculations in the type $0$ theory 
(we shall concentrate on the type 0B theory which has D3-branes,
but a similar discussion applies to the type 0A theory).
Our goal is to construct
the leading terms in the tree-level effective action $S$
for the lowest-level states in   all the four sectors:
$(NS-,NS-), (NS+,NS+),  
(R+,R+)$, and $(R-,R-)$. 
The novel features of the type $0$ theory as compared to type II  are 
the presence of the $(NS-,NS-)$ sector which contains the tachyon
(but no massless states), and the doubling of the R-R sector.
$S$ will thus involve  
the tachyon $T$, the  massless NS-NS fields $(G_{mn},B_{mn},\P)$, 
the R-R fields $(C,C_{mn},C_{mnkl}^{(+)})$ and their 
`doubles'  $(\bar C,\bar C_{mn},\bar C_{mnkl}^{(-)})$
(we shall combine $C_{mnkl}^{(+)}$ and $\bar C_{mnkl}^{(-)}$
with selfdual and anti-selfdual field strengths into a single 
field $C_{mnkl}$ with unconstrained field strength). 

The effective action has a number of interesting properties.
It is easy to see that
all the tree amplitudes which involve only the fields from the $(NS+,NS+)$
and $(R+,R+)$ (or $(NS+,NS+)$ and $(R-,R-)$) 
sectors are {\it identical}  to those in the type IIB theory
(in particular, they factorize only on states from these sectors). 
Thus, in spite of the absence of
 space-time supersymmetry, the world-sheet supersymmetry implies 
the same restrictions on the tree-level string effective action 
as in the  type II theory (e.g., the absence  of  $\a'$ corrections
to 3-point functions).  The leading $\a'$ correction to 
the second-derivative terms for the $(G_{mn},B_{mn},\P)$
fields is again of the form $\a'^3 \zeta(3) RRRR+...$, i.e.
\eqn\grvi{
S = -2 \int d^D x \sqrt G e^{-2\P} \bigg[ c_0 + 
 R + 4 (\del_n \P)^2  - {1 \ov 12}  H^2_{mnk} + O(\a'^3) \bigg] \ . }
 In this section we set $e^{\P_0}=1, \ \k= \ha$, 
so that the graviton term in the Einstein frame has the
canonical normalization.
While we  shall consider only the case of $D=10$,  for generality
we have included the central charge 
term $c_0 = - {D-10\ov \a'}$. Most of the calculations  in this section
can be repeated  for general $D$ and, in particular, the coefficients 
of the terms in the string-frame effective action below 
(apart from tachyon mass $m^2 = -{D-2 \ov 4 \a'}$) 
are independent of $D$, as they should be for consistency 
with  the world-sheet sigma-model (conformal invariance conditions) 
approach. 

The world-sheet supersymmetry also constrains 
the tachyonic sector of the action. Since $T$ is the only field
from the $(NS-,NS-)$ sector which we include in $S$,
 it is not hard to see that all
the NS-NS amplitudes involving {\it odd} powers of 
$T$, e.g.,  $\langle TTT\rangle$, 
vanish.  It is usually assumed that 
the presence of the tachyon renders the notion
of a low-energy action ill-defined. One may argue that 
there is no unambiguous derivative
expansion because, for instance, $\nabla^2 T$ may be replaced by
$m^2 T$ on shell, i.e. 
the structure of the tachyon potential is ambiguous
(see, for
instance, \refs{\bank,\tat} for discussions). 
It seems natural, however, to look for an off-shell definition 
of the effective action which satisfies the following conditions:
(i) it reproduces the string S-matrix when expanded near the 
usual tachyonic vacuum, 
and (ii) it  has other stable (non-tachyonic)  stationary points
not visible in standard string perturbation theory. 
In  the absence of some (yet unknown)  string-theoretic principle which 
unambiguously favors that  particular action 
over others which agree with the string S-matrix and have only tachyonic 
vacua, our main criterion 
is  the self-consistency of this approach:
the ability to satisfy conditions (i) and (ii) simultaneously 
is far from trivial.\foot{One may try to use the world-sheet 
sigma-model approach as  a guide in defining  the off-shell theory. 
The discussions \refs{\bank,\tat} of bosonic 
theory may not 
necessarily apply to the type $0$ string due to the presence of
world-sheet supersymmetry.  While it may look 
as if the tachyonic coupling in the sigma model
$
\int d^2\sigma d^2\theta T(X) 
\to \int d^2\sigma \psi^m \td \psi^n \del_m \del_n T 
$
is built in terms of derivatives 
of $T$ so that  the constant mode of the tachyon has no
physical meaning, 
the fact that the tachyon operator may come
in different  (not only in the $(0,0)$ but also in the $(-1,-1)$) pictures
 complicates the  situation.}
Our strategy will be to adopt the parametrization of the
tachyonic terms in the effective action
that appears most natural to us (for instance,
has the smallest number of derivatives), and to check its
consistency.

The absence of the $T^3$ interaction is an important hint that 
the type $0$ theory is more stable than the usual bosonic string:
the $T^3$ is present in the dual models  and means that 
that the leading correction to the
tachyon potential does not stabilize the theory.
On the other hand, if the leading correction is of the 
positive $T^4$ form (as discussed in section 3.2), 
then it is possible that the instability disappears in one
of the broken symmetry vacua.

Furthermore, we shall see  below that the tachyon 
is coupled to the R-R fields (sections 3.1 and 3.3).  
Thus, the curved backgrounds with non-zero R-R 
charges may lead to corrections to the effective tachyon potential and,
hopefully, to its stabilization. 

Below we shall first discuss the three-point amplitudes and then 
consider the four-tachyon amplitude
and  the  two R-R -- two tachyon amplitude. The
two graviton  -- two tachyon amplitude is discussed in Appendix A.

\subsec{Three-point amplitudes}

Since the $T^3$ amplitude vanishes, and the amplitudes involving
two tachyons and a graviton or a dilaton  are easily shown to 
correspond to the  standard covariant  kinetic term 
in the effective
action,\foot{As already mentioned above,  the 
tachyon-graviton-graviton amplitude  vanishes in type 0 theory
(in the bosonic string it does not vanish, indicating the
presence of a $TR_{mnkl}R^{mnkl}$ term in the effective action \tat).}
\eqn\term
{ 
\int d^D x \sqrt G\ e^{-2\P} ( \ha G^{mn} \del_m T \del_n T  + \ha  m^2 T^2) 
\ ,} 
 the most interesting
 calculation involves a tachyon and two
R-R massless particles. 
In  type 0B  theory the R-R fields come from the 
$(R+,R+)$ and the $(R-,R-)$ sectors. 
The $(R+,R+)$ vertex operators 
have the form
\eqn\rplus{
e^{ -\half  \p(z)  - \half  \td \p(\bz) } 
\Theta (z) \CC \G^{m_1...m_n} (1+ \Gamma_{11})
\td \Theta (\bz)\ F_{m_1...m_n} (k)\  e^{ik\cdot x(z,\bz)}
  \ , 
}
where 
the spin operators $\Theta$ and $\tilde \Theta$  are  both
$D=10$ Majorana spinors.\foot{Here we follow the 
notation of \Jbook\ and will use equations from section 12.4 there. 
$\p$ is the bosonized ghost and 
 $\CC$ is the charge conjugation matrix which satisfies
$ \CC^T = - \CC$ and anticommutes with $\G_{11}$.
Note that, in general, 
the R-R vertex operator in the sigma-model action on a sphere
has a dilaton prefactor $ e^{\P(x)}$.}   
Similarly, the $(R-,R-)$ vertex operators 
have the form
\eqn\rminus{
e^{ -\half  \p(z)  - \half  \td \p(\bz) } 
\Theta (z) \CC \G^{m_1...m_n} (1- \Gamma_{11})
\td \Theta (\bz) \ \bar F_{m_1...m_n} (k) \  e^{ik\cdot x(z,\bz)}\ . 
}
For $n=1,3$ the $n$-form field strengths are related to the
potentials $C_{n-1}$ and $\bar C_{n-1}$ through
$$F_{m_1...m_{n}}= n \del_{[m_1} C_{m_2...m_{n}]} \ , \ \ \ \ \ \
\bar F_{m_1...m_n} = n \del_{[m_1} \bar C_{m_2...m_{n}]} \ . $$
The 5-form case is special: here there is only one unconstrained field
strength $F_{m_1...m_5}$ which enters both the 
$(R+,R+)$ and the $(R-,R-)$ vertex operators,
\eqn\rfive{
e^{ -\half  \p(z)  - \half  \td \p(\bz) } 
\Theta (z) \CC \G^{m_1...m_5} (1 \pm \Gamma_{11} )
\td \Theta (\bz)\ F_{m_1...m_5}(k) \  e^{ik\cdot x(z,\bz)}
(x)   \ . 
}
The positive sign in the  projector picks out the selfdual part of the
field strength present in the $(R+,R+)$ sector, while the negative sign
picks out the anti-selfdual part present in the $(R-,R-)$ sector.

We will also need the detailed form
of the tachyon vertex operator. In the (0,0) picture it is
\eqn\taa{ k_m \psi^m (z)  k_n \td \psi^n (\bz)  e^{ik\cdot x(z,\bz)}\ . 
}
However, to saturate the ghost number conservation 
on the sphere in the  tachyon-(R-R)-(R-R) correlator, with the R-R
operators in the $(-1/2,-1/2)$ picture, the tachyon 
should be taken in the $(-1,-1)$ picture 
\eqn\taaa{ e^{ -\p(z)  -  \td \p(\bz) } e^{ik\cdot x(z,\bz)}\ . } 
Since  
\eqn\tett{\langle \Theta_\a(z_1)  \Theta_{\a'}(z_2)
\rangle = z^{-5/4}_{12} \CC_{\a\a'}  } 
is non-zero only for spinors of the {\it opposite}  chirality
($(1+\G_{11}) \CC (1-\G_{11}) = \CC$) 
we conclude that $\langle  FF T\rangle =\langle \bar F \bar F T
\rangle =0$, i.e.
the tachyon does not couple 
separately to the fields of the  $(R+,R+)$  or $(R-,R-)$ sectors (this fact
is, of course,  necessary for consistency of the type IIB theory).
At the same time,  the  $\langle  F \bar F T\rangle$ amplitude is non-vanishing 
and leads to the $F_{n} \bar F_{n} T$ terms in the effective action. 
For the 5-form field strength, the tachyon coupling is of the form
$F^{(+)}_5 F^{(-)}_5 T$, where
$F^{(\pm)}_5 =\ha ( F_5 \pm F^*_5)$ are the selfdual and anti-selfdual
parts.

The graviton-(R-R)-(R-R)  amplitude
is identical to the corresponding type II amplitude. 
Here the graviton vertex operator is  to be taken in the 
$(-1,-1)$ picture,   
\eqn\graa{ e^{ - \p(z)  - \td \p(\bz) } 
\psi^m\td \psi^n \zeta_{mn}(k)  e^{ik\cdot x(z,\bz)} 
\ . } 
Using the basic holomorphic correlators (see, e.g., \Jbook)
\eqn\duh{ \langle 
e^{ - {1\ov 2} \p(z_1)} e^{ -{1\ov 2} \p(z_2)} e^{ - \p(z_3) } \rangle 
= z_{12}^{-1/4} (z_{13} z_{23})^{-1/2}
\ , } 
\eqn\tte{
\langle \Theta_\a (z_1)  \Theta_{\a'} (z_2)  \psi^m(z_4) \rangle =
 2^{-1/2}  z_{12}^{-3/2} (z_{14}z_{24})^{-1}(\CC\G^m)_{\a\a'}\ , 
}
we get the 3-point function corresponding 
to the standard kinetic  R-R terms coupled to gravity 
(and to the  dilaton, if we switch to the Einstein frame).
As a result, the  corresponding  leading  R-R terms in the action
are 
$$ \int d^D x \sqrt G \bigg[\ \ha  |F_1|^2    + \ha  |\bar F_1|^2
+  \ha  |F_3|^2   +   \ha  |\bar F_3|^2 
 + \ha |F_5|^2   $$
\eqn\sta{
+ \  | F_1 \bar F_1| T \ + | F_3 \bar F_3| T  \ +  \ha | F_5|^2 T \ 
\bigg] \ , }
where 
$ | F_{n} \bar F_{ n}| \equiv { 1 \ov n!} 
F^{m_1...m_n}\bar F_{m_1...m_n}$. 
In the last term we used 
$$|F^{(+)}_5 F^{(-)}_5|  = \ha  | F_5|^2   $$ 
because $\ep_{10} F_5 F_5 =0$ and $\ep^2_{10} =-10!$.

The $F_1,F_3$ part of the action \sta\ may be diagonalized,  
\eqn\cond{
\int d^D x \sqrt G \bigg[ \ha (1 + T)  
|F_{1+}|^2    + \ha (1 + T)  |F_{3+}|^2  + 
 \ha (1 - T)  |F_{1-}|^2    + \ha (1 - T)  |F_{3-}|^2
\bigg] \ , }
where we introduced, as in \comb, 
$
F_{n\pm} = { 1 \ov \sqrt 2} (F_n  \pm  \bar F_n)$ (for $n=1,3$).
The  unconstrained field $F_5$ 
has both `electric' and `magnetic'
 components which  are  the 5-form counterparts of 
the  degrees of freedom in  $F_{1\pm}$ and $F_{3\pm}$. 

\subsec{The four-tachyon amplitude}

In this section we attempt to find the leading $T^4$  non-linear correction to
the tachyon potential.
The starting point is the  4-tachyon amplitude in type 0 theory. 
Taking two tachyons in the (0,0) picture and two in the $(-1,-1)$ 
picture one finds that, up to normalization, the amplitude is given by
$$A_4  =   (k_1\cdot k_2)^2  \int d^2 z  |z|^{-2 -t} |1-z|^{-2 -u} 
= 2\pi  ( 1 + \ha s)^2  
{ \Gamma (-1 -\ha s ) \Gamma ( -\ha t  )
 \Gamma ( - \ha u ) 
\over 
\Gamma ( 2 + \ha s  ) \Gamma (1 +  \ha t  ) \Gamma (1 + \ha u  ) }
\  $$
\eqn\tacc{
= - 2\pi { \Gamma (-\ha s ) \Gamma ( -\ha t  )
 \Gamma ( - \ha u ) 
\over 
\Gamma ( 1 + \ha s  ) \Gamma (1 +  \ha t  ) \Gamma (1 + \ha u  ) }
\ ,  } 
where (setting $\a'=2$) 
\eqn\moo{
s= - (k_1 + k_2)^2\ , \ \ \ \  t=-(k_1+k_4)^2\ , \ \ \ \ 
u=-(k_1 + k_3)^2\  , } 
\eqn\mon{ \ \ \ \ \ 
 \ s+t+u=4m^2 =-4\ , \ \ \  k_1+k_2+k_3+k_4=0\ , \ \ \ \ 
k^2_i= -m^2  = 1 \ . 
} 
As expected, the amplitude  is completely symmetric in $s,t,u$.\foot{Note that, 
if written in the form 
$$
 { \Gamma (1+m^2 -\ha s ) \Gamma (1+m^2 -\ha t )
 \Gamma (1+m^2 - \ha u ) 
\over 
\Gamma ( -m^2 + \ha s  ) \Gamma (-m^2 + \ha t ) \Gamma (-m^2 + \ha u  ) }
\ ,$$
this expression  gives the 4-tachyon amplitudes
in  both the bosonic  ($m^2 = - { 4 \ov \a'}$)  and 
the type 0 ($m^2 = -{2 \ov \a'}$) 
 string theories. }
It   has only massless and $m_n^2 >0$ massive poles
(in agreement with the vanishing of the 3-tachyon amplitude).
The massless  $1\ov s$ pole is found by 
expressing  $t$ and $u$ in terms of $u,s$   
and $t,s$  and sending $s$ to zero
\eqn\mass{
  { \Gamma (-\ha s ) \Gamma ( 2 + \ha u  + \ha s )
 \Gamma ( 2 + \ha t  + \ha s )
\over 
\Gamma ( 1 + \ha s  ) \Gamma (1+ \ha t  ) \Gamma (1 + \ha u  ) }
\to   - 2 { (1 + \ha t)(1+\ha u)  \ov s } +  O(s^0) \ . 
}
The residue is  proportional to $ (k_1\cdot k_3)^2$, making 
it clear that this pole is due to the  graviton/dilaton 
exchange  between the two $TT$-vertices in \term. 
Indeed,  in the Einstein frame ($G_{mn} = e^{{4\ov D-2} \P}  g_{mn}$)
the sum of \grvi\ and \term\
becomes 
$$
S = \int d^D x \sqrt g  \bigg[ -2  R  + { 8 \ov D-2}  (\del_n \P)^2 
+ \ha  (\del_n  T)^2  + \ha  m^2 e^{{4\ov D-2} \P}   T^2 + ...  \bigg]
$$
\eqn\eee{
= 
\int d^D x   \bigg[\ha  \hmn ( \D +...) \hmn  + 
 \ha \P' \D \P'   
+ \ha  T (\D + m^2) T   -\T_{mn} \hmn   + {\cal T}  \P' + ...   \bigg] \ , 
}
where  
$$\D = - \del^2\ , \ \ \ \ \ \  g_{mn} = \d_{mn} + \hmn\ ,  \ \ \ \ \ \ \
\  \P' \equiv  {4 \ov \sqrt {D-2} } \P\  ,              $$ 
and the tachyon  contributions to the sources are 
\eqn\sour{
\T_{mn} = \ha [ \del_m T \del_n T - 
\ha \d_{mn} (\del_k T \del_k T + m^2 T^2) ] \ , \ \ \ \ \ \ 
{\cal T}  =  \ha { 1 \ov \sqrt{D-2} } m^2 T^2  \ . } 
Integrating the graviton and the dilaton 
out to the leading order  we get  the 
$T^4$ exchange amplitude  which  may be written as 
a contribution to the S-matrix generating functional $\S (T^{(in)})$
\eqn\genel{
\S (T)  = \int d^D x   \bigg[
  \ha  T (\D + m^2) T    +  W (T) \bigg ] \ , 
}  $$
W  =  - \ha   \T_{mn}  \D\inv _{mn,kl} \T_{kl}    
  -\ha     {\cal T}  \D\inv {\cal T}   \ , 
$$
where $\D\inv = - \del^{-2}$ and 
$
\D\inv _{mn,kl} = (\d_{m(k} \d_{l)n} - {\textstyle { 1 \ov D-2} } \d_{mn} \d_{kl} )
 \D\inv $
is the graviton propagator in the harmonic gauge.
Evaluating the contractions in the effective
$T^4$ term in the coordinate space we find 
that all $D$-dependence 
disappears  and we are left with the 
following simple result 
$$
W 
= - \four \bigg[ 
(\del_m T \del_n T) \D\inv (\del_m T \del_n T)
- \ha (\del_n T \del_n  T  + m^2 T^2 ) 
\D\inv (\del_k T \del_k T + m^2 T^2) \bigg]
$$
\eqn\tttt{
 =\  - \four \bigg[ 
(\del_m T \del_n T) \D\inv (\del_m T \del_n T)
  +  { \textstyle{ 1 \ov 8} }   T^2 \del^2 T^2 \bigg] \ , 
}
where we  have used the on-shell condition $ \del^2 T = m^2 T$.\foot{ 
The same expression can be found in the string frame 
by 
 solving  the `beta-function'  equations  
$$R_{mn}  + 2 \nabla_m \nabla_n \P - \four \del_m T \del_n T =0 \ , \ \ 
\ \  c_0   + 2 \nabla^2 \P  - 4 (\del_n \P)^2   -\four m^2 T^2 =0 \ ,  $$
for the graviton   and dilaton in terms of 
$T$ and substituting the result back into the action. }
The corresponding expression in the momentum space 
(symmetric in $k_1,k_2,k_3,k_4$)  is  
\eqn\gen{ 
W =   \int \prod_{i=1}^4   { d^D k_i \ov (2\pi)^D }  e^{i k_i  \cdot x}
\   \W (k_1,...,k_4) \   T(k_1) T(k_2) T(k_3) T(k_4) 
\ , 
} 
\eqn\tyt{
\W_{exch} = 
-  {\textstyle { 1 \ov 12}} \bigg[  { (1+  \ha  u)(1+  \ha  t)  \ov s  }  
+  { (1+  \ha  s)(1+  \ha  t)  \ov u  }  
+  { (1+  \ha  u)(1+  \ha  s)  \ov t  }  
\bigg] \ . 
}
This is the same  massless pole  combination 
as  found in the string amplitude \mass.
Subtracting the massless exchanges from the string amplitude  we find 
that   
 the  correctly normalized  contribution to the  `contact'
(massive exchange)  
$T^4$  term in the effective action 
 is given by 
\eqn\suub{
\W_{subtr}  = {\textstyle { 1 \ov 24}} \bigg\{
 { \Gamma ( -\ha s ) \Gamma ( -\ha t ) \Gamma ( - \ha u ) 
\over 
\Gamma ( 1 + \ha s  ) \Gamma (1 + \ha t ) \Gamma (1 + \ha u  ) }
 + \bigg[ 2 { (1+  \ha  u)(1+  \ha  t)  \ov s  }  + (s,t,u \ cycle)  
  \bigg]
\bigg\} \ . } 
To  determine the  leading contact  term in this 
expression  one may first   extend it off
shell  and  then expand in powers of momenta. 
In contrast to the case when all external 
legs are  massless there  seems to be  {\it  no}  unique way of doing this,
but we suspect 
 that  the situation in type 0 theory, 
where the 4-tachyon amplitude does not have the tachyonic poles,
is better defined than in the bosonic string. 
There are at least two  expansion procedures that 
preserve symmetry between $s,t,u$:
(i) we may evaluate  \suub\  at the symmetric point 
$s=t=u = - { 4 \ov 3}$ getting a  $T^4$ term in the effective action, 
\eqn\rer{
S  = \int d^D x   \bigg[\ 
  \ha  T (\D + m^2) T    +   c_1  T^4 + ...   \bigg] \ , 
\  }
or \ (ii)  we may  set  $s=-2 - 2 k_1\cdot k_2, \ t=-2 - 2 k_1\cdot k_3,\ 
s=-2 - 2 k_1 \cdot k_4$  and expand  in powers of scalar 
products  of momenta, $k_i\cdot k_j$, treating them as independent;
that  leads to  the derivative-dependent, e.g., 
$T \del_m \del_n T \del_m T \del_n T$, terms in the effective action. 

While it is not clear if the $T^4$ correction 
may provide  a contribution to  $S(T)$ to remove the tachyon, 
we will see below that in type 0 theory there is a more viable 
mechanism for tachyon stabilization
based on the fact that the tachyon couples to the R-R fields. 
These R-R contributions to the tachyon effective potential dominate
over the possible $T^4$ terms in the limit of strong R-R
fields.

\subsec{The two R-R field  -- two tachyon amplitude }
Our aim below will be to determine the $FFTT$ 
and $\bar F \bar F TT$ terms in the effective action 
from the corresponding 4-point
 amplitudes, and also to check the coefficients of 
the  cubic terms in the action \sta\ via factorization.
For simplicity, we will first consider the R-R scalars of type $0$B
theory. Then the  generalization to other R-R particles will be quite
obvious. If the two R-R vertices  are  in  the $(-1/2,-1/2)$
picture, then
one of the  two tachyon vertex operators is  to be taken  in the 
$(-1,-1)$ picture \taa, and the other in the 
(0,0) picture \taaa.
Ordering the two R-R vertices as 1,2 and the two tachyon vertices as 3,4
and using the basic correlators \duh,\tte\ 
we find the following expression for the string amplitude
(as usual, we fix the M\"obius gauge as $z_{1,2,3}= 0,1,\infty$): 
\eqn\ami{
 k_4^m   k_4^n (\CC\G^m)_{\a\a'} (\CC\G^n)_{\b\b'} 
k_1^p k_2^q (\CC\G^p)_{\a\b} (\CC\G^q)_{\a'\b'}  \  J(s,t,u) \ .  }
\eqn\amii{
J =
\int d^2 z  |z|^{-2 -t} |1-z|^{-2 -u}  = 2\pi 
 { \Gamma (-\ha s ) \Gamma ( -\ha t  )
 \Gamma ( - \ha u ) 
\over 
\Gamma ( 1 + \ha s  ) \Gamma (1 +  \ha t  ) \Gamma (1 + \ha u  ) }
 } 
  is the same ratio of $\G$-functions   
as in the  final expression 
for the 4-tachyon amplitude in  \tacc,
except for the  fact that now 
\eqn\now{
s+t+u=-2 \ ,  \ \ \ \ \ \   k^2_1=k^2_2=0\ , \ \ \  \ \ k^2_3=k^2_4=1 \ . }
In the kinematic factor (which is non-vanishing 
only if the two R-R vertices are from the same sector)
we have used the fact that each  R-R  scalar 
vertex   contains  the field strength $F_m= \del_m C$ 
and thus an  extra factor of momentum.   
Since  $\CC \G^m  = - (\G^m)^T \CC, \ \  \CC^2 =1$, 
one finds that  the kinematic factor reduces to 
\eqn\yty{ \Tr ( \G_m \G_p \G_n \G_q)  k_4^m   k_4^n k_1^p k_2^q
= 32( 2 k_1\cdot k_4 k_2\cdot k_4  -  k_1\cdot k_2 k_4\cdot k_4) =
 16 (-1+ tu ) \ .  }
The string  amplitude has massless poles  in $s,t$, and $u$ channels.
The $s$-channel pole ($\sim  {tu-1 \ov s}$) 
is due to the graviton/dilaton exchange
between the R-R and the tachyon `stress tensor' vertices, while
the $t,u$ poles are due to the R-R  field exchange 
between the two $F_1\bar F_1 T$ vertices in \sta.
The normalisation can be fixed by reproducing the 
 $s$-channel pole from field theory.  
We start with  the effective action
parametrized by two constants $a,b$  ($C$ and $\C$ are the R-R scalars)
$$
S  = \int d^D x  \sqrt G   \bigg[
  e^{-2\P} ( -2R - 8 \del^m\P  \del_m \P + 
\ha  \del^m T \del_m T  + \ha m^2 T^2 )
  $$ \eqn\seea{
   + \   \ha  ( \del^m  C \del_m  C +  
\del^m  \C  \del_m  \C ) ( 1 +\  b\  T^2 + ... ) 
 + \  a\  \del^m  C \del_m  \C  \ T  +  ...  \bigg] \ , 
}
or, in the Einstein frame, 
$$
S  = \int d^D x  \sqrt g   \bigg[
  -2R  + \ha  \del^m\P' \del_m  \P' + \ha  \del^m T \del_m T  + \ha
e^{{1  \ov \sqrt{D-2} } \P' }   m^2 T^2 
  $$ 
\eqn\yyt{
   +  \  \ha e^{{ 1 \ov 2} \sqrt{D-2} \P'  }  
( \del^m  C \del_m  C +  \del^m  \C  \del_m  \C ) ( 1 +\ b \ T^2 + ... ) 
 +\  a  \ e^{ { 1 \ov 2}\sqrt{D-2}\P'  }  \del^m  C \del_m  \C  T  +  ...
 \bigg] \ .
}
Calculating the graviton and dilaton 
exchange  $CC-TT$   amplitude we find
that, as in $TT-TT$ case \tttt, 
all $D$-dependence cancels out 
and  (after use of the on-shell conditions for the legs)
we are left with  the  following  simple 
result  for $W$ in \genel
$$
W_{grav}= - \four 
(\del_m C \del_n C) \D\inv (\del_m T \del_n T - 
\ha \d_{mn} \del_l T \del_l T) $$ \eqn\cctt{
 = \ - \four \bigg[ 
(\del_m C \del_n C) \D\inv (\del_m T \del_n T)
  +  { \textstyle{ 1 \ov 4} }   C^2 \del_l T \del_l T  \bigg] \ . 
}
In the momentum space (symmetrizing over $t,u$) 
we get for the analogue of $\W$ in \gen
\eqn\trew{ 
\W^{(s)}_{field} = \W_{grav} =   \trto 
\left [ { (1+u)^2  + (1+t)^2  \ov s}  -  ( 2 + s) \right ] 
 = \sixt { 1-tu \ov s} 
\  . } 
This is the same $s$-channel pole we got in the string amplitude.
This fixes the overall coefficient in the string 
amplitude  which thus 
corresponds to the following  $CCTT$ term 
in the generating functional for the S-matrix
(the kernel below  is to be multiplied by $ C(k_1)C(k_2)
T(k_3) T(k_4)$ and integrated over momenta  as in \gen)
\eqn\str{ 
\W_{string} =  \trto  (- 1 +tu)  { \Gamma (-\ha s ) \Gamma ( -\ha t  )
 \Gamma ( - \ha u ) 
\over 
\Gamma ( 1 + \ha s  ) \Gamma (1 +  \ha t  ) \Gamma (1 + \ha u  ) } \ . 
} 
The  $\C$ exchange in field theory \yyt\ 
 gives the following contribution to the $CCTT$ amplitude 
\eqn\uuu{
W_{form} =-  \ha a^2 \del_m  C \del_m T  \D\inv \del_n   C \del_n T  
 \  \to \  \ 
 \W_{form} = \sixt a^2 \left [    
{ (1+ u)^2 \ov u}  +  { (1+ t)^2 \ov t} \right ] 
\ . }
The leading $1\ov t$ and $1\ov u$ poles  are as  
in  the string amplitude  \str\  if 
\eqn\lil{  a^2 = 1 \ , }
 which  is  the same value   of $a$ as in \sta.
The contact $\del C\del C TT$ vertex   in \yyt\ 
combined with the $\bar C$-exchange part  \uuu\ gives 
the following  contribution to   the  field-theoretic 
$CCTT$  amplitude $\W$ (which is to be added to the $s$-channel exchange
 \trew) 
\eqn\fff{
\W^{(t,u)}_{field}=  \W_{form} + \W_{cont} = 
\sixt  \bigg[    { (1+ u)^2 \ov u}  +  { (1+ t)^2 \ov t}   + 4 b s  \bigg] \ . 
}
To fix the constant $b$ let us compare 
 \fff\  with the string-theory expression 
for the sum of the $t$ and $u$ channel poles. Rewriting \str\ as 
\eqn\rert{
\W_{string} =   \sixt  { ( 1 -tu) (t+u) \ov tu}  P(t,u) \ , \ \ \ \
P\equiv 
 { \Gamma (1 + \ha t + \ha u  ) 
\Gamma ( 1-\ha t  )
 \Gamma ( 1- \ha u ) 
\over 
\Gamma ( 1 -\ha u -\ha t  ) \Gamma (1 +  \ha t  ) \Gamma (1 + \ha u  ) }
\ ,  } 
and taking  the limit $t,u\to 0$, we get 
$P=1 +\  c_1  ut(u+t) + O(t^5,u^5)$,
where $c_1 = [{ d^3\ov dz^3 } \log \G (z) ]_{z=1}$.
Thus
\eqn\uuo{
\W_{string}(t,u\to 0) =  \sixt  ( 1 -tu)  ( { 1 \ov t} + {1 \ov u}) + O(t^2,u^2)
\ , } 
and comparison with the field-theory result \fff\ gives
\eqn\beb{
b= \ha 
\ . } 
It is worth emphasizing  the difference between 
the $TTTT$ amplitude  discussed above 
and the present case: 
there the field-theory exchanges were reproducing 
the leading terms in the expansion of the string-theory amplitude; 
here in the $t,u$ channels this 
happens only if the extra contact $CCTT$ and $\bar C \bar C TT$ 
terms are added to the effective action. 

Similarly, one finds 
that the effective action \sta\
contains the terms 
 $ \four (|F_3|^2 + |\bar F_3|^2  + |F_5|^2) T^2$. 
We will be particularly interested in the
$F_5$ background which is created by the threebrane solution.
The $F_5$-dependent terms in the action that we have determined are
(cf. \cond) 
\eqn\dac{ \int  d^D x \sqrt G \ 
 \ha  (1 + T + \ha T^2)  |F_5|^2
 \ .  }
It is instructive to sketch an explicit derivation of these terms.
Let us start with the field theory \seea\ 
and replace the $C,\bar C$  part by 
$ \ha  (1 + aT + bT^2)  |F_5|^2$. 
 The sum of the
 leading-order  graviton  and $C_4$ exchanges, and 
the contact  contributions 
 to the  $F_5F_5TT$ part of the field-theory generating functional
 is found to be (cf. \cctt, \uuu)
$$
W_{field}= W_{grav} + W_{form} + W_{cont}
$$  
$$
=\  {\te{ 1\ov 8 \cdot 5!} }(10 F_{m klpq} F_{nklpq} -  
\d_{mn} F_{sklpq} F_{sklpq})
\del^{-2} (\del_m T \del_n T) 
$$ 
\eqn\fiee{
 +\ 
{\te{ 1\ov 2 \cdot 4!}} a^2 (F_{m klpq} \del_m T)
 \del^{-2} (F_{n klpq} \del_n T)
+  {\te{ 1\ov 2 \cdot 5!}} b\  F_{m klpq} F_{mklpq} T T \  . }
The string amplitude computed  using  the  non-chiral R-R vertex operator
\rfive, i.e. 

\noindent $e^{ -\half  \p(z)  - \half  \td \p(\bz) } 
\Theta  \CC \G^{m_1...m_5} 
\td \Theta\ F_{m_1...m_5}
(k) e^{ik\cdot x}$, contains instead of \yty\ the
 following kinematic factor
 (here it is useful to keep the external field strength as a single object 
without separating the momentum factor in it):
\eqn\ytyi{ \Tr ( \G_m \G^{p_1...p_5} \G_n \G^{q_1...q_5}) \ 
 F_{p_1...p_5}(k_1)  F_{q_1...q_5}(k_2)\   k_4^m   k_4^n   \ . }
The  total amplitude is  similar to the one in \str,  
  $$
\W_{string} =  {\te{ 1 \ov 16 \cdot 5!}}
 \big[ 10 F_{mklpq}(k_1)  F_{n klpq}(k_2)  k_4^m   k_4^n 
 -  F_{mklpq}(k_1)  F_{m klpq}(k_2) \  k_4\cdot k_4 \big]  $$
\eqn\stri{ \times\ 
 { \Gamma (-\ha s ) \Gamma ( -\ha t  )
 \Gamma ( - \ha u ) 
\over 
\Gamma ( 1 + \ha s  ) \Gamma (1 +  \ha t  ) \Gamma (1 + \ha u  ) } \  . 
} 
Extending this kinematic factor  off shell in the  most 
obvious  way gives the  following expression for the sum of the 
$s$-channel  and the $t,u$ channel poles of the amplitude \stri\
(in the coordinate space representation)
 $$ 
W_{string} =  - {\te{ 1 \ov 8 \cdot 5!}}
\big[ 10 (F_{mklpq}  F_{n klpq} )   \del^{-2} ( T \del_m \del_n T  ) 
 -  (F_{nklpq}  F_{n klpq})\del^{-2}  (T \del^2 T)   \big]
$$
\eqn\strip{ 
 - \ {\te{ 1 \ov 4 \cdot 5!}}
\big[ 10 (F_{mklpq} T)  \del^{-2}  (F_{n klpq} \del_m \del_n T  ) 
 -  (F_{nklpq} T) \del^{-2} ( F_{n klpq} \del^2 T)   \big] \ . }
Now we integrate by parts and use the $F_5$ equation of motion,
$\del_m F_{mklpq} =0$, and  the  Bianchi identity, 
$F_{mklpq} (\del_n F_{mklpq} - 5 \del_m F_{nklpq} ) =0$. 
As in the R-R scalar case, we then find 
that the $s$-channel pole in \stri\ is in agreement with 
the graviton exchange in \fiee, and the sum of the $t,u$ poles
is in agreement with  the sum of the $C_4$ exchange and contact term in 
\fiee,  iff $a^2=1, \ b=\ha$.  This establishes explicitly
that the $F_5$-dependent terms in the effective action
are given by \dac. 

While it may seem that, since the string  kinematic  factor 
in \stri,\strip\ involves only derivatives of the tachyon field, 
the effective action should also involve $\del T$, 
it is  worth stressing that the contact $F_5F_5 TT$  term  
with no derivatives on $T$ comes out of the expression \strip\ upon 
rearranging it in the form \fiee. We believe, therefore, that 
our off-shell definition of the  action  as given in \dac\ 
is a natural one.

In the next section we will discuss some implications of the  action \dac\
for threebrane solutions in type 0B theory.

\newsec{The threebrane near-horizon solution }

In this section we discuss the threebrane solution. More specifically,
we focus on its near-horizon limit, but it is
precisely this region that is expected to be dual to the
gauge theory \refs{\jthroat}. In the type IIB case,
this throat region has the geometry of $AdS_5\times S^5$ with a constant
selfdual 5-form flux. The fact that the dilaton is constant indicates
the scale-independence of the gauge coupling, which is in accord with
the vanishing of the beta function in the dual ${\cal N}=4$
supersymmetric $SU(N)$ gauge theory.

The $SU(N)$ gauge theory that we are considering is {\it non}-supersymmetric:
it is related to the ${\cal N}=4$ theory through removing all the fermionic
partners. This theory is asymptotically free in the ultraviolet and is
expected to be confining in the infrared.
In fact, it may be in the same universality class as the pure glue
theory, since the 6 adjoint scalars $X^i$ may
become massive in the absence of supersymmetry.
In the gravity (effective action) approximation we can
study this theory reliably only for large bare coupling.
While it is hard to
predict precisely what should happen in this limit, it would be
strange indeed if we found that its gravity dual is still 
$AdS_5\times S^5$.\foot{In fact, we will argue that for large
bare coupling the gravitational background has a tachyonic instability.}
This would correspond to a line of fixed points that
we can essentially rule out on the gauge theory side.
Luckily, we are able to demonstrate that $AdS_5\times S^5$ is not a
solution of the type $0$B theory with an electric 5-form flux. 
The mechanism for breaking
of conformal invariance is quite subtle and involves tachyon
condensation.

One lesson that we learned from the analysis of elementary
type $0$B D3-branes is that they can carry two types of charges,
electric and magnetic. If we wish to construct a non-supersymmetric
gauge theory on coincident D3-branes, 
we may stack D3-branes of one type:
say, the positively charged electric D3-branes. Thus, the dual
gravity description will carry $N$ units of electric 5-form flux.
For such a 5-form background $|F_5|^2$ does not vanish, and from the
action in \dac\ we find a shift of the tachyon (mass)$^2$, and also
an effective linear term for the tachyon.
If the 5-form field were taken to be selfdual,
then $|F_5|^2$ would vanish and none of the
interesting effects that we are discussing would occur.\foot{This
is the field that surrounds a self-dual 3-brane, and it is quite
easy to see that in the type 0B theory there exists an 
$AdS_5\times S^5$ background with self-dual 5-form flux and
the vanishing tachyon. 
The bulk tachyon simply decouples from the
type 0B selfdual 3-brane.}
Thus, the doubling of the R-R sector in type $0$B theory 
appears to be crucial for shifting the tachyon.

The string frame field equations  
following from \grvi, \dac\ can be put into the form 
(cf. footnote 8; \ $\k_{10} = \four$)
\eqn\fii{ c_0+ 2 \nabla^2 \P  - 4 \na^n \P \na_n 
\P  - \four m^2 T^2 =0 \ ,  }
$$ 
R_{mn}  + 2 \nabla_m \nabla_n \P-  \four \na_m T \na_n T   $$
 \eqn\fiii{ - \ 
{\textstyle{1 \ov 4 \cdot 4!}} e^{2 \P}  f(T) 
(F_{m klpq} F_{n}^{\ klpq}
 - {\textstyle { 1\ov 10}} G_{mn} F_{s klpq} F^{s klpq}) =0 \ , }
\eqn\fiy{ (-\nabla^2 + 2 \nabla^n \P \nabla_n   + m^2 ) T + 
  {\textstyle{1 \ov 2 \cdot 5!}}  e^{2 \P} f'(T) F_{s klpq} F^{s klpq} =0 \ , \  }
\eqn\more{ \na_m [ f(T) F^{mnkpq} ] =0 \ ,}
where
$$  
f(T) \equiv 1 + T + \ha T^2  \ . 
$$
The central charge deficit term $c_0= { 10-D \ov \a'} $ vanishes
in the present case.
Note that \fii\  
does not depend on the 5-form terms in the action.
This is a consequence of the Weyl-invariance of the $|F_5|^2$ term
in 10 dimensions, i.e. the fact that it drops out of the trace of eq. \fiii. 

Equation \fii\ can be rewritten in the following useful form:
\foot{If there are corrections to
the tachyon potential in the absence of R-R fields, then
$\ha m^2 T^2$ in $M^2$ should be replaced by $V(T)= \ha m^2 T^2 + c_1 T^4 +
\ldots$, i.e. $M^2(T) = -\ha V(T)$.
This will not seriously affect the R-R stabilization 
mechanism of the tachyon.}
\eqn\diil{
- \nabla^2 e^{-2\P}  +    M^2 e^{-2\P} =0  \ , \ \  \ \ \ \ \ \ \ \ \
M^2 \equiv -  \four m^2 T^2  = {\textstyle{  1 \ov 2 \a'}} T^2 \ . }
Remarkably, $ e^{-2\P} $ is {\it not} tachyonic precisely because $T$ is. 
This equation can always be `integrated once':
indeed, the equation for $\P$ can always be written in the first-order
form as follows from eliminating $\nabla^2 \P$
from \fii\ using the trace of \fiii:
\eqn\iii{ 
4  \nabla^n \P \nabla_n \P  =  -R 
+ \four \nabla^n T \nabla_n T     - \four m^2 T^2 \ .  }
The first-order character of this equation with respect to the dilaton 
 seems  consistent with the    interpretation of the  radial $e^\P$ 
equation as  the RG evolution equation for ${g_{\rm YM}^2}$. 

It is quite easy to analyze the case where $|F_5|^2$ is so large that
it dominates over the quadratic action \term. 
Assuming $e^{2\P}
|F_5|^2\gg |m^2|$, a  particular solution of the  tachyon  equation is 
\eqn\pac{T=T_{vac}=\const \ , \ \ \ \ \  \ \ \ f'(T_{vac}) =0  \ . }
Then $M^2$ in the equation \diil\  for $\P$ becomes constant, and 
also \iii\  reduces to 
\eqn\iiii{ 
| \nabla_n \P |^2  =  \four ( M^2  -R )  \ .  }
In the background where the tachyon
field acquires vacuum expectation value $T_{vac} = -1$
the dilaton thus has a source term coming from the
quadratic tachyon action (the $|F_5|^2$ terms are Weyl invariant and do
not generate a dilaton source). 
 Thus, in the Einstein frame the dilaton
equation is
\eqn\uiu{ \nabla^2 \P =-{\te{ 1\over 4\alpha'} } 
e^{{1\ov 2}\P} T_{vac}^2
\ .
}  
This seems to imply that
the dilaton decreases toward larger distance from the brane, i.e.
the coupling indeed appears to decrease in the UV region, as expected
from the asymptotic freedom of the gauge theory. 
The running of the dilaton  means that the conformal invariance is
lost, and $AdS_5\times S^5$ is not a solution. The fact that the {\it tachyon
condensation} is the mechanism for the {\it breaking of 4-d conformal
invariance} provides support for our scenario.

It is interesting to note that, under the assumption that
an approximate solution has $T=\const$,
the equations \fii,\fiii,\more,\diil\
become formally the same as in the  {\it non-critical} 
string theory without a tachyon condensate, but with 
the `effective'  central charge deficit 
$c_{eff} = M^2 = {\textstyle{  1 \ov 2 \a'}} T^2$. 
The non-zero value of $M^2$ in \diil\ sets a scale for  
the gradients of the fields and, in the Einstein frame \uiu, 
plays the role of the  coefficient of the exponential dilaton potential 
(making it hard to find the solution analytically, as we explain below). 

Let us  look for  the electrically charged 3-brane background 
using  the  following  
ansatz  for the metric and the R-R field
 \eqn\asi{
ds^2 =  d\t^2 + e^{2\l(\t)} (-dt^2 + dx_i dx_i)  + e^{2\n(\t)} d\Omega^2_5 \ , }
\eqn\tre{
C_{0123} = A(\t) \ , \ \ \ \ \  \ \ \ \ F_{0123 \t} = A'(\t)
\ , } 
where $\t$ is related to the radial direction transverse to the 3-brane
($i=1,2,3$).
All  the fields including the tachyon are, in general, 
 functions of 
$\t$. Substituting the solution of the equation for $C_4$,  
\eqn\see{
 A' = 2 Q  e^{  4\l -  5 \n} f^{-1}(T) \ , \ \ \ \ \ \ \ \  Q=\const \ , 
}
into the remaining equations \fii,\fiii,\fiy, 
we find that, as usual \gibb, they can be  derived  from  an 
effective action $S(\l,\n,\P,T)$
 for a  mechanical  system  with a potential.\foot{One gets
 the wrong sign  of the  `electric' $Q^2$-term in $S$ 
 if one substitutes \see\ directly
into the action --  when varying the action over $\l,\nu$ 
one is not taking  into account the variation of $A$.
The `mechanical' equations are to be supplemented by the 
`zero-energy' constraint  
following from  $G_{\t\t}$ variation of the original action.}

The resulting action and equations in the general case of $T\neq \const$
 are  given in Appendix B, 
while  here we  shall simply assume, as suggested above, 
that for large 3-brane charge $Q$  one 
may ignore   the tachyonic mass term in the 
equation for $T$   so that there  should exist a self-consistent solution
with $T=\const, \ f'(T)=0$.  This should be true 
 for large $Q$ in the near-horizon region provided 
\eqn\xxx{
   Q^2   e^{2\P -  10\n}f^{-1} (T) \  \gg\  \eight |m^2| T^2 
\ . }
Under the assumption that $T=T_{vac} =-1$ 
is an approximate solution we get  $f(T_{vac}) = \ha $, \ $M^2 
= {\textstyle{  1 \ov 2 \a'}} T^2_{vac} =  {\textstyle{  1 \ov 2 \a'}}$. 
Introducing the  new `time' parameter $\r$  and the fields 
$\xi,\eta$ 
\eqn\nnew{
d\r =e^{2 \P - 4 \l - 5 \n } d\t  \ , \ \ \ \ \ \ 
\xi =  \P    - 4\l   \ , \ \ \ \ \
\eta =   \P - 2  \l -  2 \n     \ , }
we find   that our problem is described by the following 
Toda-like mechanical system (dot denotes $\r$-derivative)
\eqn\daas{  
S=  \int d\r \bigg[ \ha \dot  \P^2  + \ha \dot \xi^2 - 5 \dot \eta^2  
   - V(\P,\xi,\eta) \bigg] \ , 
}
\eqn\iuio{
V =   M^2 e^{{1 \ov 2}\P + {1 \ov 2}  \xi   - 
5 \eta }   + 20 e^{-4\eta }   -  2 Q^2   e^{-2\xi}  \ 
. } 
The  three terms in the potential  have a clear interpretation:
the first  
originates from the `effective central charge' or tachyon mass term,
the second  represents  the curvature of $S^5$, and the third
is produced by the electric  R-R charge.  
The zero-energy constraint that supplements 
the second-order equations following from 
\daas, 
\eqn\zerr{
\ha \dot  \P^2  + \ha \dot \xi^2 - 5 \dot \eta^2  
+  V(\P,\xi,\eta)  = 0 \ ,  }
is  equivalent to \iiii\ (with the
second derivatives of the metric components
in $R$ eliminated using  their equations of motion). 
 When $M^2=0$ 
the dilaton has no potential and $\xi$ and $\eta$ 
are described by decoupled Liouville-type actions. 
As a result, one finds the  electric  analogue 
of the standard  R-R 3-brane solution \hsdgt:
\eqn\soo{ \P=\P_0 \ , \ \ \ \ \ 
e^\xi  = e^{\P_0}  + 2 Q  \r \equiv  e^{\P_0} H \ , \  \ \ \ 
e^\eta = 2 \sqrt \r \ , \ \ \ \
\r =  { e^{2\P_0}  \ov 4r^4}  \ ,  }
which implies
$$
d\t^2 = H^{1/2} dr^2\ ,\quad  e^{2\l} = H^{-1/2}\ , \quad 
e^{2\n} = H^{1/2} r^2\ , \quad 
 H=1 + { e^{\P_0} Q \ov 2 r^4}\ .
$$
When $M^2\not=0$  all  the three fields are coupled 
and  the  exact analytic solution does not seem possible. 
This increase  in  complexity is a general feature of the black hole 
solutions in the presence of a dilaton potential \refs{\gibb,\wilt}.\foot{To get a feeling for the complexity of this system let us 
make a simplifying assumption that 
for large $Q$ the  curvature of $S^5$ is supposed 
to be small so that the second term in $V$ \iuio\ may be ignored. 
Then  one finds that 
$ \P = \eta + q \t + p  , \    q,p=\const$ and 
$\ddot \xi = \ha  M^2 e^{{1 \ov 2}\P + {1 \ov 2}  \xi   - 5 \eta } 
  + 4 Q^2   e^{-2\xi} $, \  
$\ddot \eta = - \ha  M^2 e^{{1 \ov 2}\P + {1 \ov 2}  \xi   - 5 \eta }
$
plus the constraint \zerr. Even this simplified system of equations does
not appear to be exactly solvable.  }
It should be sufficient, however, to analyze the 
solution qualitatively in the near-horizon region $r\to 0$
(or, equivalently, the asymptotic 
`long-time' region  $\r \to \infty$).  
 We shall not  attempt to do 
this here (for some potentially relevant discussions
see \wilt).

Although the complete  near-horizon solution is not known, we would like to
estimate the range of parameters for which the tachyon instability is
removed. For this purpose we will assume that, just as for
$AdS_5\times S^5$, the overall scale of the metric is $\sqrt{N g_s}$.
The condition for the effective (mass)$^2$ of $T$ to be positive 
(which is, essentially, equivalent to \xxx) is 
\eqn\stabcond{ {1\over 4} g_s^2 |F_5|^2 > {1\over \alpha'}
\ .
}
Since $F_5\sim N$, while the inverse metric is of order $1/\sqrt{N
g_s}$, \stabcond\ implies
\eqn\newstab{
{1\over \sqrt{N g_s}} > O(1)
\ .}
In the dual field theory $N g_s$ is the bare `t Hooft coupling, which
we need to send to zero in order to achieve the continuum limit.
It is satisfying that for small $N g_s$ the type $0$B background indeed
appears to be stable. However, if the stability criterion \newstab\
is satisfied, then there are significant corrections to the gravity
approximation. Thus, in order to study such RR-charged
backgrounds reliably, we really need to find the corresponding 
conformal $\sigma$-models.

\newsec{Discussion}
In this paper, following up on the idea of \AP, we studied possible
connections between type $0$ strings and non-supersymmetric
gauge theories. It turns out that non-supersymmetric, yet tachyon-free,
field theories are naturally realized as the world volume theories
on coincident D-branes of the type $0$ theory. By analogy with
the recently discovered AdS/CFT duality, we attempt to give dual
descriptions of these large $N$ gauge theories in terms of the R-R-charged 
$p$-brane classical backgrounds. An immediate problem is the presence
of a tachyon in the closed string spectrum. However, based on our analysis
of the tachyon effective action in the presence of R-R field strength,
we argue that the tachyon may condense, while its effective (mass)$^2$
may become shifted to a positive value. This provides an appealing
mechanism for curing the tachyonic instability. Our analysis of the
near-horizon threebrane geometry shows that the tachyon instability
is indeed removed, while the condensation breaks the conformal
invariance, in agreement with the fact that the dual gauge theory
is not conformal.

While we feel that we have constructed an appealing scenario
for a new generalization of the AdS/CFT duality, much work remains
to be done in checking its consistency. The only truly convincing check
would be the construction of a conformal field theory with the R-R and
tachyon backgrounds, but unfortunately we are still far from this goal.

\newsec{Acknowledgements}
We are grateful to A.M. Polyakov for very useful discussions.
A.A.T. is also grateful to H. Liu for  discussions. 
The work  of I.R.K.  
was supported in part by the NSF
grant PHY-9802484 and 
by the James S. McDonnell
Foundation Grant No. 91-48.  
The  work  of A.A.T. was supported in part
by PPARC, the European
Commission TMR programme grant ERBFMRX-CT96-0045
and  the INTAS grant No.96-538.



\appendix{A}{ The two graviton -- two tachyon amplitude }

Let us first make some general comments on which terms we 
should expect in the graviton-tachyon sector 
of the  effective Lagrangian. On general grounds, one may have (in the Einstein frame,  up to factors 
 of dilaton and terms with derivatives of dilaton)
\eqn\gene{
L= a_1\a'   R^{mn} \nabla_m 
T \nabla_n T  + a_2\a'   R \nabla^k T \nabla_k T  + 
a_3  R T^2   +  T TTT \ terms \ ,  }
where $TTTT$ stands for $T^4$ as well as terms with derivatives of 
the tachyon field. 
Using equations of motion
\eqn\ioi{R_{mn} =\ha (  \T_{mn} - { 1 \ov D-2} g_{mn} \T_{kk} )  \ , 
\ \ \ \   R=  - { 1\ov D-2}  \T_{kk}\  , \ \ \ \  \nabla^m \T_{mn} =0 \ , }
where $\T_{mn}$ was defined in \sour\ 
and we have taken into account $\nabla^2 T= m^2 T$,  
we can thus transform  the $R TT$ terms in \gene\ 
into  the $ TTTT$ terms.  That means 
that such $RTT$ terms can be redefined away and 
give a vanishing contribution to 
the $hhTT$ amplitude\foot{Note  that while 
the  $R TT$ terms do not contribute to $hTT$ 3-point function, they could still contribute to the 
4-point amplitude $hhTT$  
(e.g., $R = - \four \del_k h_{mn} \del_k h_{mn} + ...$).
However, the contribution  of such contact vertices cancels against 
that of the exchange diagram  obtained 
by pairing  $hTT$ vertex from the $RTT$ with $hhh$ vertex from $R$ term.}
 in particular. Their contribution to the 4-tachyon amplitude
is equivalent to that of  the contact $TTTT$ terms in the action.
Since the coefficients of  such $RTT$ 
terms are thus ambiguous, it 
 is natural to define the effective action  by setting them equal to zero.
At the same time, it is natural to keep the 
 $FFTT, \bar F \bar F TT$  and  $TTTT$  terms    in the action.

Our aim will be to 
compare the string-theory and field-theory
expressions for the $hhTT$ amplitude.
Let us  first compute the field-theory amplitude. 
We shall  consider only at the  $h_{mn} h_{mn}$ part of  the amplitude 
where the polarization tensors of the two 
gravitons are contracted onto each other (not on momenta).
The relevant  $h^2$ and $h^3$ terms in  the  expansion of the  Einstein term 
in the action  near flat space are
($g_{mn} = \d_{mn} + h_{mn}$) 
  $$
- 2 \sqrt g R= 
\ha  \del_k  h \del_k h  
-  \ha  h_{kl} \del_k h_{mn}  \del_l h_{mn}  
$$ \eqn\expa{+\  \four h_{ll} \del_k h_{mn}  \del_k h_{mn}
-  h_{mn}  \del_k h_{mn} \del_l h_{kl} 
+  h_{mn}  \del_k h_{mn}  \del_k h_{ll}  
- h_{kl} \del_n h_{mk}  \del_n h_{ml} \ . 
}
We have included the  last 
`cyclic contraction' term $h_{kl} \del_n h_{mk}  \del_n h_{ml}$
since it also contributes to the
 $h_{mn} h_{mn} TT$  amplitude.
After integration by parts and dropping terms 
where $\del^2$ acts on one of the contracted  (external)
$h_{mn}$ legs we get 
\eqn\exa{
- 2 \sqrt g R= 
\ha  (\del_k  h_{mn}  \del_k h_{mn}+ ...)     - h_{kl} \TT_{kl}
 - h_{kl} \del_n h_{mk}  \del_n h_{ml}
   \ , \ \ \
} $$
\TT_{kl} \equiv    h_{mn}  \del_k \del_l h_{mn} 
+ \ha \del_k h_{mn} \del_l h_{mn} 
- {\textstyle{3 \ov 4}} \d_{kl} 
 \del_k h_{mn}  \del_k h_{mn}  
  \ , \ \ \ \  \del^k \TT_{kl} =0 \ . 
$$
The relevant 
 terms in the tachyonic action  are (we  ignore
 the dilaton dependence as the dilaton 
exchanges do not contribute to the $hhTT$ amplitude)
\eqn\poi{
\ha  (\del_k T \del_k T + m^2 T^2 ) (1 - \four h_{mn} h_{mn})
 - h_{mn } \T_{mn}  
\ , } 
where the stress tensor $\T_{mn}$ is the same as  in \sour. 
Integrating out the graviton  connecting the   $TTh$
 and    $hhh$ vertices and adding also the contact $hhTT$
 term we find
for the $hhTT$ term in the resulting 
generating functional (see \genel) 
\eqn\rew{
W= W_{cont} + W_{exch} \ , \ \ \  \ \ \ \ 
W_{cont} =  - \eight \del_k h_{mn} \del_k h_{mn} T T 
\ , } 
   \eqn\exx{ 
W_{exch} = - (\TT_{mn}  - \del_p h_{qm} \del_p h_{qn})
 \D\inv_{mn,kl} \T_{kl}   }
 $$
= - \four h_{mn} \del_k \del_l h_{mn} \del^{-2}  \del_k T \del_l T 
+ \eight \del_k h_{mn} \del_k h_{mn} T^2
+ {\textstyle{ 1 \ov 4(D-2)}}  m^2 T^2
 h_{mn}  h_{mn} $$   $$ - \   {\textstyle{ 1 \ov 4(D-2)}} 
 m^2 T^2 h_{mn}  h_{mn}   \ , $$
where the last term is the contribution 
of the `cyclic contraction' term.\foot{We omit  other 
terms with different index contraction patterns coming from the 
`cyclic' term.  Note  also that the relative 
 signs of  $\TT$ and the `cyclic' term are different in \exa\  and in \exx:
there are 3 possible contractions of $h_{pq}$ in the `cyclic' 
term, with two  having opposite sign to that of the third one.}
The  total  result  is  very simple  
\eqn\uoi{
W= - \four ( h_{mn} \del_k \del_l h_{mn} ) \del^{-2} ( \del_k T \del_l T )
\ ,  } 
were we have  used the 
 mass shell condition for the  tachyon  and graviton legs 
to transform  
$h_{mn} h_{mn} (\del_k T \del_k T + m^2 T^2)$ into 
$\del_k \hmn \del_k \hmn  T^2$. 
In the momentum space we get  (this should be multiplied by the 
product of the graviton  and tachyon    factors 
$\zeta_{mn}(k_1) \zeta_{mn}(k_2) T(k_3) T(k_4)$ as in \gen) 
\eqn\hhtt{
\W= - \sixt  \   { (1+t) (1+u) \ov s}  \ , 
}
where we used that 
$$
k^2_{1,2} =0\ , \ \ \ \ \     k^2_{3,4} =-m^2 =1\ , \ \ \ \  \ \ 
s+t+u=-2 \ . 
$$
The string   $hhTT$   amplitude 
is easily computed 
in type 0 theory, but for simplicity we shall  again 
consider only its part where  the two graviton 
polarization tensors 
are contracted onto  each other.
The  amplitude is found by taking the  
two gravitons (particles 1 and 2) in  the 
$(-1,-1)$ picture \graa\  and the two tachyons (particles 3 and 4) 
in the (0,0) picture \taa.
As in \tacc\ the momentum factor $(k_3\cdot 
k_4)^2 = ( 1+ \ha s)^2$
comes from the contraction of fermions in the 
two tachyonic vertices, and the rest of the amplitude 
is given by  $\int d^2 z  |z|^{-1-t} |1-z|^{-1-u}$, 
i.e., up to normalization, 
\eqn\ito{
\ \ \ \ A_4=   ( 1 + \ha s)^2    
{ \Gamma (-1-\ha s ) \Gamma ( \ha - \ha t  )
 \Gamma (\ha - \ha u ) 
\over 
\Gamma ( 2 + \ha s  ) 
\Gamma (\ha + \ha t   ) \Gamma (\ha + \ha u  ) }
=  \ { 2 \ov s}\ { \Gamma (1-\ha s ) \Gamma ( \ha - \ha t  )
 \Gamma (\ha - \ha u ) 
\over 
\Gamma ( 1 + \ha s  ) 
\Gamma (\ha + \ha t   ) \Gamma (\ha + \ha u  ) } \ . 
}
The amplitude is $t-u$ symmetric, 
has  massless 
pole in the $s$-channel and $no$ tachyonic poles.\foot{Though the 
 two $\del T\del T h$ vertices could be, in principle, 
 connected by the 
tachyon propagator, 
 this does not produce the  graviton polarization tensor factor
we are considering.}
The pole $A_4(s\to 0) \to \ha  { (1+u)(1+t)  \ov s}  + O(s^0)$
is in perfect agreement with the field theory result \hhtt,
implying that the normalized expression for the 
string amplitude with massless exchange subtracted is  given by
(cf. \suub) 
\eqn\eqe{
\W_{subtr}=  - \sixt
{ (1+u)(1+t)  \ov s} \bigg[ { \Gamma (1-\ha s ) \Gamma ( \ha + \ha t + \ha s   )
 \Gamma (\ha + \ha u + \ha s ) 
\over 
\Gamma ( 1 + \ha s  ) 
\Gamma (\ha + \ha t   ) \Gamma (\ha + \ha u  ) }  -1 \bigg]  \ . 
}

\appendix{B}{Classical equations for electric p-brane 
backgrounds}

Here we shall discuss the general system of equations 
corresponding to the ansatz \asi,\tre\ and $T=T(\t)$. For generality we 
will replace the 5-dimensional sphere in \asi\ by $k$-dimensional 
one, the number 3 of brane spatial dimensions   by $p$ 
and the rank of the R-R potential  by $n=p+1$. Thus, the case 
discussed in   Section 4 
is $k=5, n=4$. The parametrisation \asi\
is similar to the one used  in the study 
of cosmological solutions with a non-trivial dilaton \tva,
with $\t$ palying the role of (Euclidean) time. 
The corresponding effective action for the fields 
$\l,\n,A,T$ and the redefined  dilaton 
\eqn\dio{ \vp \equiv 2 \P - n \l - k \n } 
  which follows from \grvi,\dac\ is 
$$S=-  \int d\t\ \bigg(  e^{-\vp} \bigg[  c_0  +  k(k-1)   e^{-2\n}
  -  n  \l'^2 - k \n'^2 
 + \vp'^2   - \four T'^2 - \four m^2 T^2     \bigg]   $$
\eqn\acu{
 +\    \four  e^{ - n \l + k \n} f(T) A'^2  \bigg)
\ .  }
 Substituting the solution of the equation for the R-R potential 
\eqn\sie{
 A' = 2 Q  e^{  n\l -  k \n} f^{-1}(T) \ , \ \ \ \ \ \ \ \  Q=\const \ 
}
into the remaining equations \fii,\fiii,\fiy\ 
we find that they become (prime denotes the $\t$-derivative)
\eqn\onn{
   n  \l'^2 + k \n'^2  -  \vp'^2  +   \four T'^2   + \V = 0    \ , \  }
\eqn\onna{
 \l'' - \vp' \l'  =   - {{ 1 \ov 2n}}   {\pa \V \over \pa \l} \ ,\ \ \ \ 
 \ \ \ \ \ 
 \n''  - \vp' \n'   =    -   {{ 1 \ov 2k}}{\pa   \V \over \pa \n} \ ,
  }
\eqn\onnb{
 \vp'' - n \l'^2 -  k \n'^2  = { 1 \ov 2}  {\pa \V \over \pa \vp}    \ ,  }
\eqn\onnt{
 T'' - \vp' T'  =   -  2 {\pa \V \over \pa   T} \ .  }
These equations can be  derived from  the following action:
\eqn\au{
S=   \int d\t\    e^{-\vp} \bigg[  n  \l'^2 + k  \n'^2 
 -  \vp'^2    +   \four T'^2 
 - \V  (\l,\n,\vp,T) \bigg] \ , } 
\eqn\abi{
 \V \equiv   c_0  + k(k-1) e^{-2\n} 
 - \four  m^2 T^2  -    Q^2   e^{\vp +   n \l -  k\n} f^{-1} (T)  \ ,  
}
provided one adds also the 
 zero-energy constraint \onn.

The action \au\  may be transformed into a simpler form by 
making  redefinitions similar to \nnew.
Specifying to the case of our interest, $n=4, \ k=5$, 
we thus find the following action  which 
generalizes \daas,\iuio\ to the case of non-constant tachyon:
\eqn\eas{  
S=  \int d\r \bigg[ \ha \dot  \P^2  + \ha \dot \xi^2   
- 5  \dot \eta^2 + \four \dot T^2
   - V(\P,\xi,\eta,T) \bigg] \ , 
}
\eqn\eio{
V = e^{-2\vp} \V=   (c_0  - \four  m^2 T^2)  
e^{{1 \ov 2}\P + {1 \ov 2}  \xi   - 5 \eta }
   + 20   e^{-4\eta }   -   Q^2   e^{-2\xi}  f^{-1} (T) \
. } 
The 
3-brane solution corresponding to this action will, in general, 
involve $T$ changing with $\r$. 
As we discussed in section 4, for large $Q$
it is natural to expect that  for large $\r$
(or near the horizon) the tachyon will asymptotically approach a 
constant value. It would be interesting to try to 
confirm this in a  more rigorous way.

\vfill\eject
\listrefs
\end